\documentclass[12pt]{iopart}

\usepackage{graphicx}
\usepackage{iopams}

\usepackage{epsfig}
\usepackage{graphicx}
\usepackage{cite}
\usepackage{ams}
\usepackage{amsfonts}
\usepackage{amssymb}
\usepackage{array}
\usepackage{setspace}

\newcommand\eg{{\textsl{e.g.\,}}}
\def\BE{\begin{equation}}
\def\EE{\end{equation}}
\def\BEA{\begin{eqnarray}}
\def\EEA{\end{eqnarray}}
\newcommand{\cut}[1]{{}}

\begin{document}

\title{An Efficient CDMA Decoder for Correlated Information Sources}
\author{Hadar Efraim$^{1}$, Nadav Yacov$^{1}$, Ori Shental$^{2}$ and Ido Kanter$^{1}$}

\address{$^1$  Minerva Center and Department of Physics, Bar-Ilan University, Ramat-Gan 52900, Israel}
\address{$^2$ Center for Magnetic Recording Research (CMRR), University of
California, San Diego (UCSD), 9500 Gilman Drive, La Jolla, CA 92093,
USA}

\begin{abstract}
We consider the detection of correlated information sources in the
ubiquitous Code-Division Multiple-Access (CDMA) scheme. We propose a
message-passing based scheme for detecting correlated sources
directly, with no need for source coding. The detection is done
simultaneously over a block of transmitted binary symbols (word).
Simulation results are provided demonstrating a substantial
improvement in bit-error-rate in comparison with the unmodified
detector and the alternative of source compression. The robustness
of the error-performance improvement is shown under practical model
settings, including wrong estimation of the generating Markov
transition matrix and finite-length spreading codes.
\end{abstract}

\maketitle

\section{Introduction}
Code division multiple access (CDMA) is a core technology of today's
wireless communication employing data transmission between multiple
terminals and a single base station. Usually in the uncoded CDMA
literature~\cite{BookVerdu}, the binary information source,
modulated for transmission over the channel is assumed to be taken
from an \emph{unbiased} identically independently distributed
(i.i.d.) random process.

In practice, however, a substantial level of redundancy can often be
observed in real-life uncoded sources (e.g., uncompressed binary
images) which can be viewed as a (global) bias in the generating
Bernoulli distribution or as correlations between the binary
symbols. These correlations can be considered as a \emph{local} bias
on a certain symbol in the generating Bernoulli distribution. In
such cases, a binary source encoding is used. A source encoder is
said to be optimal if it can eliminate all source redundancies and
eventually generate unbiased outputs. However, most existing
practical source encoders, which are typically fixed-length
encoders, are sub-optimal. Besides the extensive complexity of the
encoder, its output still contains a certain level of bias or
correlations, which can be further exploited in the transceiver
design.

As for coded systems, it has been shown~\cite{ShamaiVerdu} that the
empirical distribution of any 'good' error-correcting code converges
to the channel's capacity-achieving input distribution. A 'good'
code is a code approaching capacity with asymptotically vanishing
probability of error. Hence, well-coded information sources, in the
common case of binary-input additive white Gaussian noise (BI-AWGN)
CDMA channel, must be unbiased, as the capacity-achieving input
distribution of the BI-AWGN channel is Bernoulli 1/2~\cite{Tanaka}.
However, bias (global and local) can be found in practical coded
CDMA systems in which not so 'good' codes are being employed, \eg,
systematic Turbo codes~\cite{Berrou,Zhu1,Zhu2}, for which the
systematic component of the code entails bias.

A different approach than source compression for handling CDMA with
biased sources was lately introduced~\cite{Efraim}. In the presented
scheme the source bias $m$ was assumed to be estimated by the
receiver and used for modifying both the naive single-user
matched-filter (SUMF) and state-of-the-art CDMA multiuser detectors (MUD).
That modification outperformed the alternative of applying source
coding by detecting modulated biased sources directly, with no need
of source encoding.

A substantial level of redundancy can also be viewed as correlations
between the binary symbols with no (global) bias. In this case the
aforementioned scheme is useless. Still the correlations, which
contain local bias, can be used for an improved detection.

Therefore, tuning CDMA detection for correlated sources is of major
importance in both coded and uncoded settings. In this paper, we
examine the commonly-used random spreading scheme, which lends
itself to analysis and well describes CDMA with long signature
sequences. We suggest a scheme for handling CDMA with correlated
sources, rather than using source compression.

We propose a scheme in which the source correlations are assumed to
be estimated by the receiver, or reported periodically to the
receiver via an auxiliary low-rate channel. The source correlations
are used for evaluating the applied local bias, $m$, on each
transmitted binary symbol, followed by adding a correction factor of
the form $\tanh^{-1}{m}$ to the multiuser CDMA
detector~\cite{Efraim}, recently introduced by
Kabashima~\cite{Kabashima}. The proposed scheme outperforms the
alternative of applying source coding throughout a wide range of
practical correlation values. Also the derived scheme is shown to
yield an improved error performance not only for a large system
limit (the limit where the number of users and spreading codes tend
to infinity but with a fixed ratio, which is defined as the system
load $\beta$), but also for finite-length spreading codes. Note that
the proposed scheme is not optimal, yet heuristically presents a
good error performance and convergence rate. The BER performance may
be affected due to the finite size effect of the transmitted word of
each user, where the correlation length is comparable. Also it may
be influenced  in case where the applied local bias leads to a
non-optimal fixed point of Kabashima's algorithm.

\section{CDMA with correlated information sources}
Consider a $K$-user synchronous direct-sequence binary phase
shift-keying (DS/BPSK) CDMA system employing random binary spreading
codes of $N$ chips over additive white Gaussian noise (AWGN)
channel. The received signal of such a system can be expressed as
\BE \label{eq_ymu} y_{\mu}=\frac{1}{\sqrt{N}}\sum _{k=1}^{K}s_{\mu
k}b_{k}+n_{\mu}, \EE where $s_{\mu k}=\pm1$ ($\mu=1,\ldots,N$,
$k=1,\ldots,K$) are the binary spreading chips being independently
and equiprobably chosen. The deterministic chip waveform is assumed
to be of unit energy; $n_{\mu}$ is an AWGN sample taken from the
Gaussian distribution ${\mathcal{N}}(0,\sigma^2)$; $b_{k}$ is the
(possibly coded) information source binary symbol transmitted by the
$k$'th user and is modeled by a Markov process, generated by a
Markov transition matrix. In this paper we demonstrate our scheme
for information source binary symbols which are modeled by a 2-state
(2S) Markov process, generated by the following Markov transition
matrix \BE \label{eq_transfer_matrix}\mathbf{T}_{ab}= \left(
\begin{array}{cc}
T_{-1 -1} & T_{-1 1} \\
T_{1 -1} & T_{11} \\
\end{array} \right)
\EE where $T_{a b}$ ($a,b=\pm1$) is the probability of transmitting
a binary symbol $b$ following the transmission of a binary symbol
$a$. We assume that each user sends a word, each is assembled from
$L$ symbols. 

We work under two scenarios regarding the Markov transition matrix. In the first scenario the matrix is assumed to be known to the receiver as side information, for example, after reported via an auxiliary low-rate feedback channel from the receiver. The second scenario refers to the case where there is no side information channel. In this case the estimation of the Markov parameters ($T_{a b}$) can be done in each iteration of the decoding process according to the tentative value of the decoded words of all the users. According to the probabilities of each decoded binary symbol to be $\pm1$, the probabilities of decoding a binary symbol $b$ after $a$ (i.e, $T_{ab}$) can easily be evaluated.

We assume a prefect power-control mechanism yielding unit energy transmissions. Also we assume a
situation where $N$ and $K$ are large, yet the system load factor
$\beta=K/N$ is kept finite.

The goal of the MUD is to simultaneously detect binary symbols
$b_{1}, b_{2},...,b_{K}$ after receiving the signals $y_{1},
y_{2},...,y_{N}$. The Bayesian approach offers a useful framework
for this. Assuming that the binary signals are independently
generated from the unbiased distribution, the posterior distribution
from the received signals is provided as
\begin{equation}P(\textbf{b}|\textbf{y})=\frac{\prod_{\mu=1}^{N}P(y_{\mu}|\textbf{b})}{\sum_{\textbf{b}}\prod_{\mu=1}^{N}P(y_{\mu}|\textbf{b})}\end{equation}
where
\begin{equation}P(y_{\mu}|\textbf{b})=\frac{1}{\sqrt{2 \pi \sigma^2}} exp\left[ \frac{-1}{2 \sigma^2}(y_{\mu}-\Delta_{\mu})^2 \right]\end{equation}
and
$\Delta_{\mu}=\frac{1}{\sqrt{N}} \sum_{k=1}^{K}s_{\mu k} b_{k}$.

Recently, Kabashima~\cite{Kabashima} has introduced a tractable
iterative CDMA MUD which is based on the celebrated belief
propagation algorithm (BP,~\cite{BookPearl,BookOpperSaad}). This
novel algorithm exhibits considerably faster convergence than
conventional multistage detection~\cite{Andrews} without increasing
computational cost significantly. It is considered to provide a
nearly-optimal detection when the spreading factor $N$ is large and
the noise level is known. Similarly to multistage detection, at each
iteration cycle $t$ this detector computes tentative soft decisions
$\eta_{k}^{t}$ for each user transmission, of the form \BE\label{eq
eta} \eta_{k}^{t}=\tanh(h_{k}^{t}). \EE

The parameters $\eta_{k}^{t}$ and $h_{k}^{t}$ are coupled and being iteratively computed using the
following recipe \BEA\label{eq_U1}
U_{k}^{t}&=&A^{t}\sum_{l=1}^{K}W_{kl}\eta_{l}^{t}+A^{t}\beta(1-Q^{t})U_{k}^{t-1},\\
h_{k}^{t+1}&=&R^{t}h_{k}^{0}-U_{k}^{t}+A^{t}\eta_{k}^{t}(1-Q^{t})U_{\mu}^{t-1},\\
R^{t}&=&A^{t}+A^{t}\beta (1-Q^{t})R^{t-1},\label{eq_U2} \EEA where $W_{kl}\triangleq\sum_{\mu=1}^{N}s_{\mu k}s_{\mu l}/N$, $Q^{t}\triangleq\sum_{k=1}^{K}(\eta_{k}^{t})^{2}/K$, $A^{t}\triangleq(\sigma^{2}+\beta(1-Q^{t}))^{-1}$ and tentative hard-decisions are taken by
$\hat{b}_{k}^{t}=\textrm{sign}(\eta_{k}^{t})$. Producing $\hat{b}_{k}^{t}\equiv \hat{b}_{k}^{t+1}, \forall k,$ serves as the convergence criterion.

\section{Improving detection for correlated sources}
In order to exploit all the possible knowledge for the detection of
each symbol, let all the words of all the users ($b_{k}^{l},
b_{k}^{l+1},...,b_{k+1}^{l}, b_{k+1}^{l+1}...$, where $k=1..K$ and
$l=1..L$) be transmitted altogether to the receiver (total amount of
$L \times K$ binary symbols). As a result, in the decoding process,
the decoder has a maximum knowledge of the correlations of
$b_{k}^{l}$ (in our case with $b_{k}^{l-1}$ and $b_{k}^{l+1}$). As
described in (\ref{eq_ymu}) the transmitted binary symbols are
modulated over AWGN channel. In this case, of transmitting all the
words of all the users, we reformulate (\ref{eq_ymu}) to have the
new form of the received signal \BE \label{eq_ymut}
y_{\mu}^{l}=\frac{1}{\sqrt{N}}\sum _{k=1}^{K}s_{\mu
k}b_{k}^{l}+n_{\mu}. \EE

With the received signals, the decoder applies a SUMF according to
\BE\label{eq_hk} h_{k}^{l}=\frac{1}{\sqrt{N}} \sum_{\mu=1}^{N}
y_{\mu}^{l} s_{\mu k}. \EE Using the output of the SUMF,
$h_{k}^{l}$, the decoder evaluates the probabilities of each symbol
to be $\hat{b}_{k}^{l}=1$ or $\hat{b}_{k}^{l}=-1$ according to
 \BE\label{eq_prob1a} q_{k}^{l}(\hat{b}_{k}^{l}=1)=\frac{1+\tanh(h_{k}^{l})}{2}
\EE and
\BE\label{eq_prob1b}q_{k}^{l}(\hat{b}_{k}^{l}=-1)=\frac{1-\tanh(h_{k}^{l})}{2}.\EE

Following eq. (\ref{eq_prob1a}) and (\ref{eq_prob1b}) and using the
Markov transition matrix, $T_{ab}$, the detector evaluates the
applied local bias on each bit using the following recipe:
\BE\label{eq_prob2a} p_{k}^{l}(\hat{b}_{k}^{l}=1)=\sum_{a,b=\pm1}
q_{k}^{l-1}(a)\cdot T_{a1}\cdot q_{k}^{l+1}(b)\cdot T_{1b},\EE
\BE\label{eq_prob2b} p_{k}^{l}(\hat{b}_{k}^{l}=-1)=\sum_{a,b=\pm1}
q_{k}^{l-1}(a)\cdot T_{a-1}\cdot q_{k}^{l+1}(b)\cdot T_{-1b},\EE and

\BE\label{eq_mk} m_{k}^{l}=2\cdot
\frac{p_{k}^{l}(\hat{b}_{k}^{l}=1)}{p_{k}^{l}(\hat{b}_{k}^{l}=1)+p_{k}^{l}(\hat{b}_{k}^{l}=-1)}-1.\EE

We can now incorporate a correction factor for error-performance
improvement, based on the local bias (\ref{eq_mk}), to the SUMF
output (\ref{eq_hk}) and finally have a binary decision
\BE\label{eq_MF_biased}
\hat{b}_{k}=\textrm{sign}(h_{k}^{l}+\xi_{k}^{l}), \EE where
$\textrm{sign}(\cdot)$ is the hard-decision signum function and
$\xi_{k}^{l}=(\beta+\sigma^{2})\tanh^{-1}{m_{k}^{l}}$
~\cite{Efraim}.

We can also employ the tractable iterative CDMA MUD (\ref{eq eta}-\ref{eq_U2}).
In order to adapt the MUD for correlated sources and to the case of
detecting all the words of all the users simultaneously, we
reformulate (\ref{eq eta}) to have the new form

\BE\label{eq eta_correlation}
\eta_{kt}^{l}=\tanh(h_{kt}^{l}+\xi_{kt}^{l}), \EE where
$\xi_{kt}^{l}=\tanh^{-1}{m_{kt}^{l}}$ is a correction factor for
error-performance improvement being incorporated within the
detection algorithm~\cite{Efraim}.

After each iteration of the MUD, an evaluation of the probabilities
of each symbol to be $\hat{b}_{k}^{l}=1$ or $\hat{b}_{k}^{l}=-1$ is
assessed according to (\ref{eq_prob1a},\ref{eq_prob1b}), followed by
a calculation of the local bias which is applied on each symbol
(\ref{eq_prob2a},\ref{eq_prob2b},\ref{eq_mk}) and finally by another
iteration of the MUD with the correction factor (\ref{eq
eta_correlation}).

\section{Results and discussions}
In this section, simulation results of the proposed scheme for
correlated sources are presented. Unless stated otherwise, all the
results are obtained for load $\beta=0.8$ and $\sigma=0.8$ while
simulation results are averaged over sufficiently large ensemble of
$2000$ computer-simulated randomly spread AWGN CDMA samples with
long spreading factor $N=1000$.

\begin{figure}[t!]
\begin{center}
\begin{tabular}{cc}
{\scalebox{1}[1]{\includegraphics[width=0.5\textwidth]{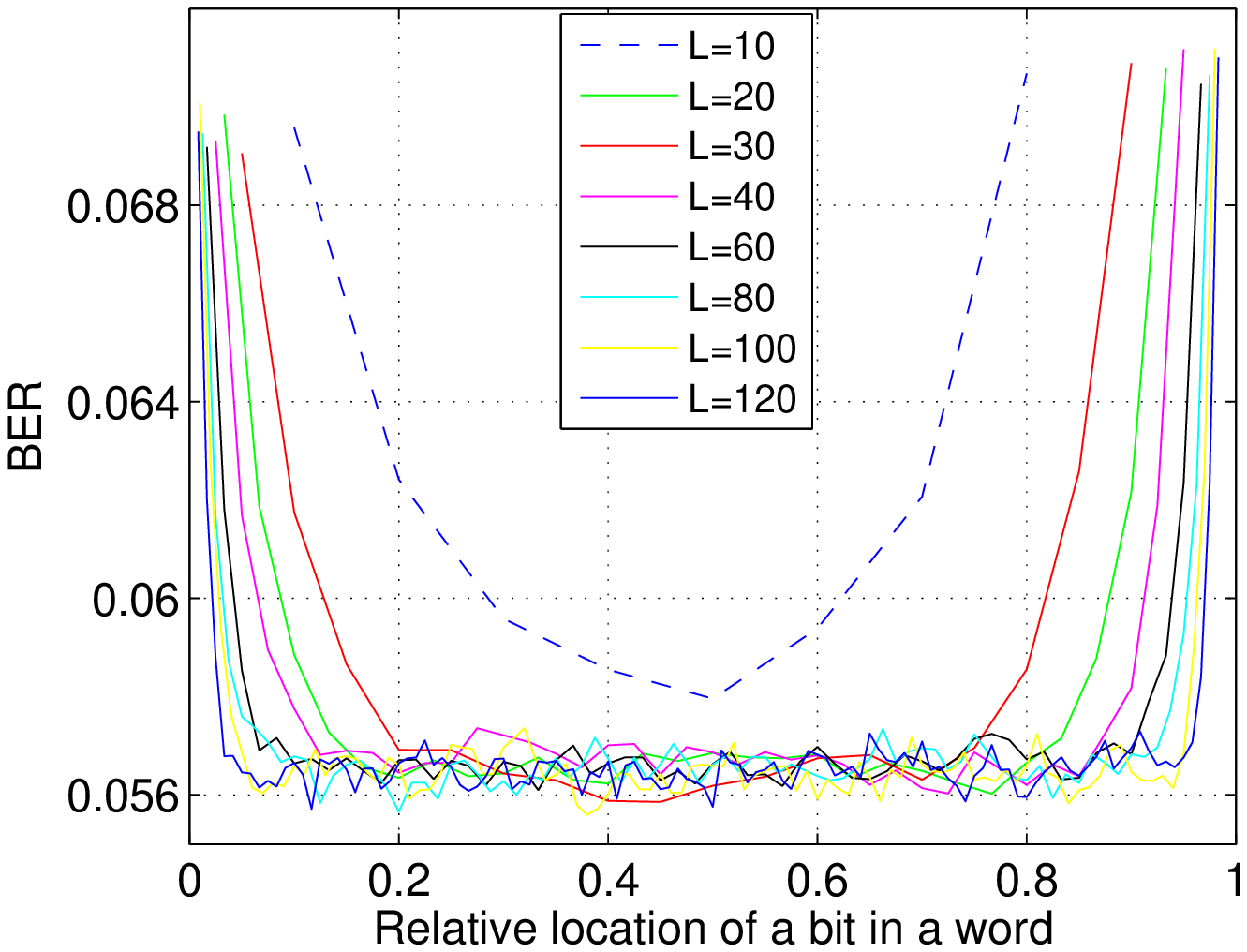}}}
&
{\scalebox{1}[1]{\includegraphics[width=0.5\textwidth]{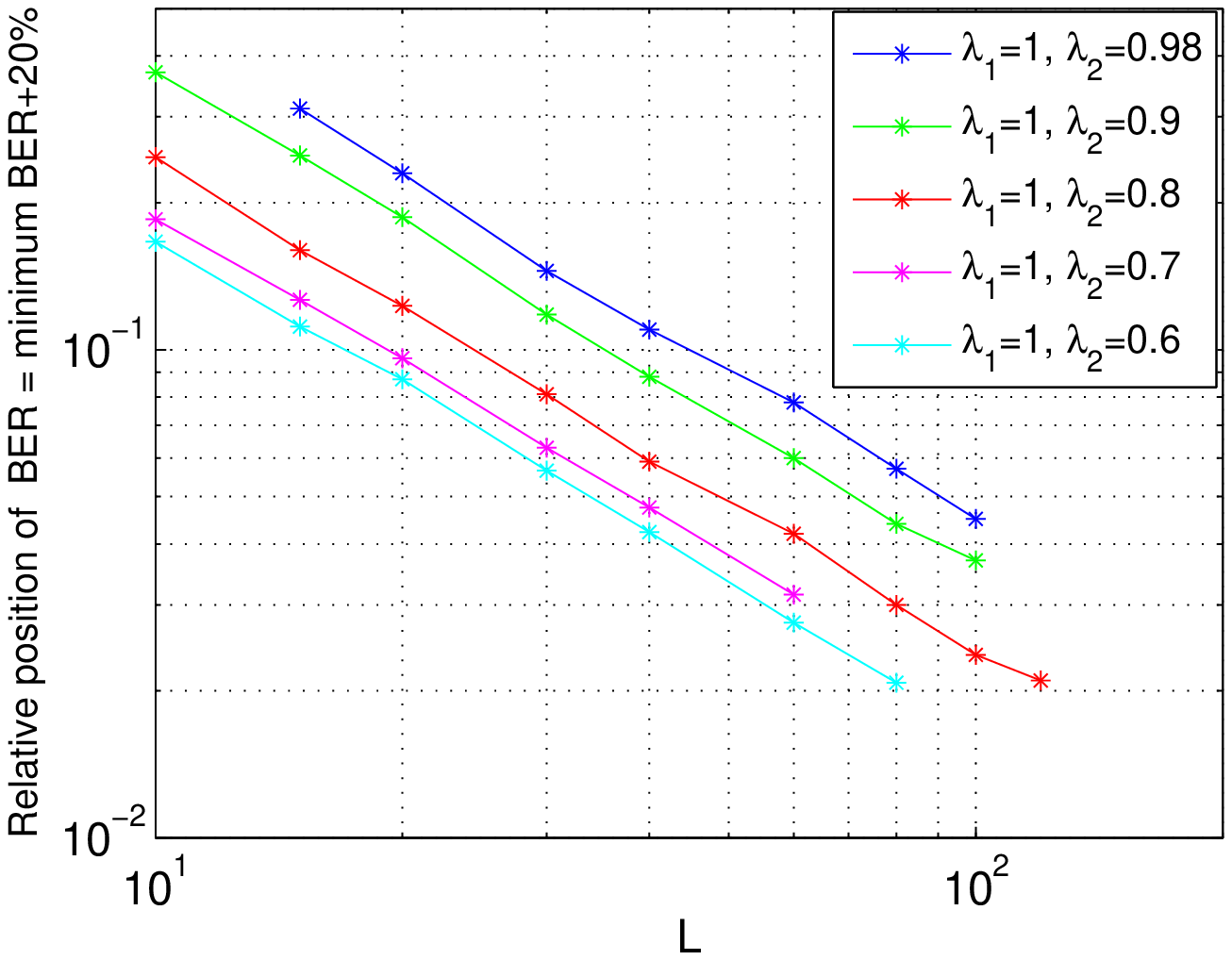}}}
\\(a) & (b)\\
{\scalebox{1}[1]{\includegraphics[width=0.5\textwidth]{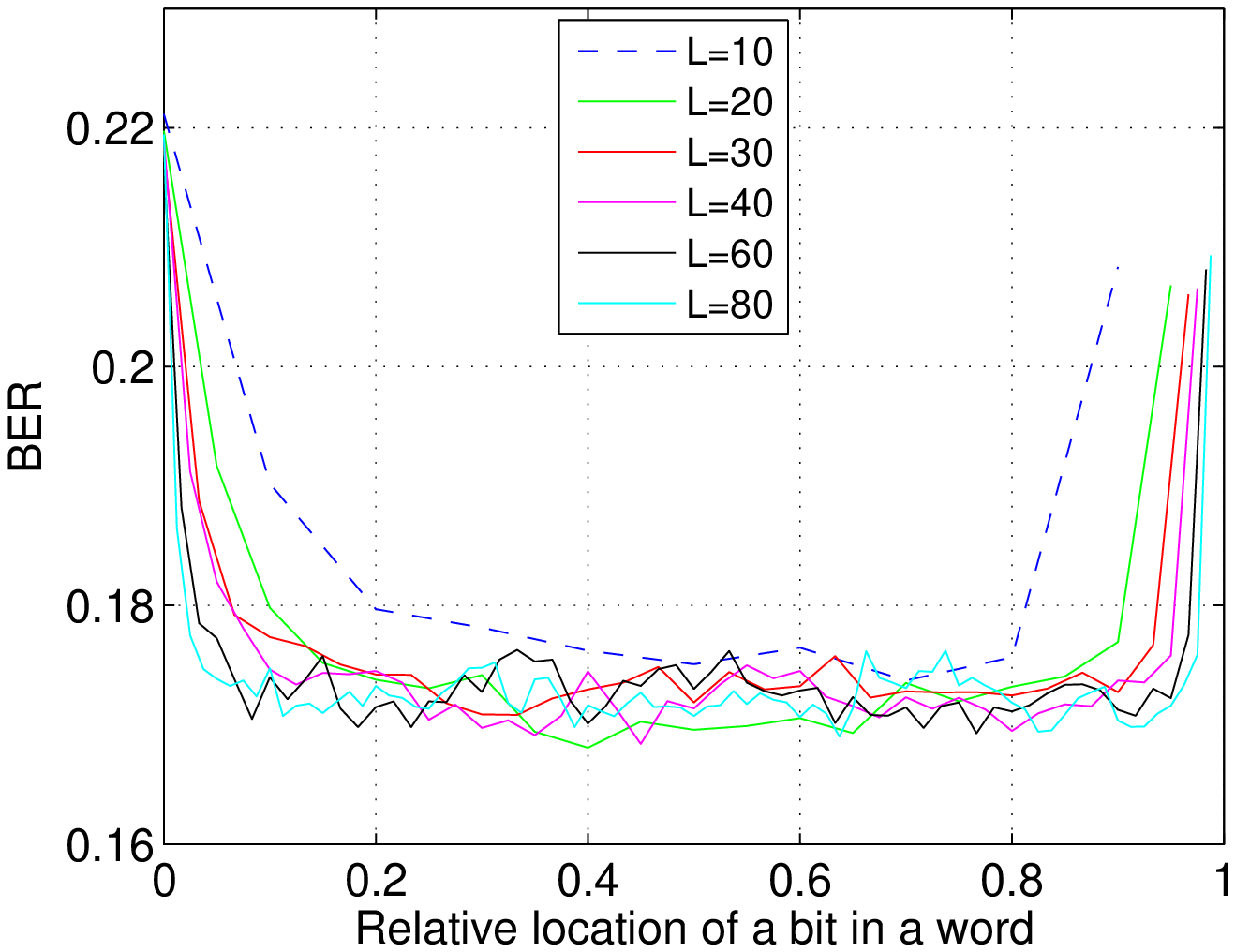}}}
\\ (c) \\
\end{tabular}
\end{center}
\vspace{-0.5cm} \caption{(a) Bit error rate (BER) vs. the relative
location of a bit in a word for $N=1000$, $\beta=0.8$, $\sigma=0.8$,
Markov transition matrix with eigenvalues $\lambda_{1}=1$ and
$\lambda_{2}=0.8$ and for several length of words $L=10, 20, 30,
 40, 60, 80, 100, 120$. Simulation results are averaged over $2000$ samples. (b) The relative position in the detected word where the BER is
20\% above the minimum BER vs. the length of the detected word, $L$,
for $N=1000$, $\beta=0.8$, $\sigma=0.8$ and for several Markov
transition matrices. Simulation results are averaged over $2000$
samples. (c) SUMF BER vs. the relative location of a bit in a word
for $N=1000$, $\beta=0.8$, $\sigma=0.8$, Markov transition matrix
with eigenvalues $\lambda_{1}=1$ and $\lambda_{2}=0.8$ and for
several length of words $L=10, 20, 30,
 40, 60, 80$. Simulation results are averaged over $2000$ samples.}\label{fig_ber_detection}
\end{figure}

Figure~\ref{fig_ber_detection}-(a) displays the bit error rate (BER)
of the proposed scheme (\ref{eq_prob1a}-\ref{eq_mk}, \ref{eq
eta_correlation}) as a function of the relative location of a
certain binary symbol in the detected word. The relative location of a certain bit is its location compared to the beginning of the detected word, divided by the length of the detected word. For example, in the case where the length of the detected word is $L=30$, the relative location of the $3rd$ bit is $0.1$.  The results were
obtained for a Markov transition matrix with eigenvalues
$\lambda_{1}=1$ and $\lambda_{2}=0.8$ and for several lengthes of
words $L=10, 20, 30, 40, 60, 80,1 00, 120$. Results indicate that
the portion of the word that has a substantial and stable
improvement of the BER increases with the size of the word. This
tendency can be explained by using the correlation length of the
Markov process,
$\xi=\frac{1}{\ln{\frac{\lambda_{1}}{\lambda_{2}}}}$. The effect of
a wrong detected bit over other bits in the word vanishes for a
length grater than $\xi$. For example, the correlation length of the
aforementioned generating Markov transition matrix is $\xi=4.48$.
Thus, for a detected word with length of $L=120$ more than $95 \%$
of the bits are with (considerably) low and stable BER, while on the
other hand for a detected word with length of $L=10$, we can hardly
see a saturation of the BER. Similar results were also obtained for
small ($N=25$) CDMA systems and for other Markov transition
matrices.

An interesting feature of the results in
Figure~\ref{fig_ber_detection}-(a) is the convergence to saturation
of the BER as a function of the length of the detected word.
Figure~\ref{fig_ber_detection}-(a) shows that the convergence to the
asymptotic BER decreases with the size of the word.

We measured the behavior of the convergence to saturation of the BER
as a function of the length of the word, $L$.
Figure~\ref{fig_ber_detection}-(b) displays the behavior of the
convergence to saturation for correlated sources, generated by
several Markov transition matrices. We calculated the convergence to
saturation as the relative position in the detected word where the
BER (which is displayed in Figure~\ref{fig_ber_detection}-(a)) is
$20 \%$ above the level of the minimum BER.

Figure~\ref{fig_ber_detection}-(b) indicates that for all correlated
sources, generated by different Markov transition matrices, the
convergence to saturation of the BER behaves like $L^{-1}$. Similar
results were obtained for CDMA systems with small spreading, $N=25$,
and with other Markov transition matrices.

Figure~\ref{fig_ber_detection}-(c) displays the BER of the proposed
scheme while using (only) the popular and naive SUMF
(\ref{eq_prob1a}-\ref{eq_MF_biased}) as a function of the relative
location of a certain binary symbol in the detected word. The same
behavior as demonstrated in Figure~\ref{fig_ber_detection}-(a),(b)
is also presented here.

\begin{figure}
\begin{center}
{\scalebox{1}[1]{\includegraphics[width=0.67\textwidth]{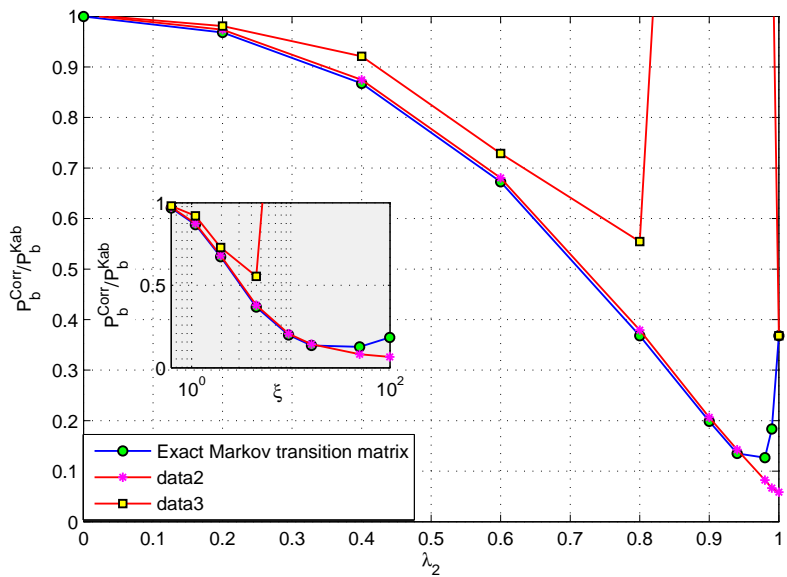}}}
\end{center}
\vspace{-0.5cm} \caption{Normalized BER, $P_{b}^{Corr}/P_{b}^{Kab}$,
as a function of the eigenvalues of the Markov transition matrices
($\lambda_{1}=1$ and $\lambda_{2}$) for $N=1000$, $\beta=0.8$,
$\sigma=0.8$ and $L=100$, averaged over 2000 samples. Also drawn are
the results of the BER under a mismatch of $\pm 10 \%$ of element
$T_{-1-1}$. Normalized BER as a function of the correlation length $\xi$ is drawn in the inset.}\label{fig_norm_ber}
\end{figure}

Figure~\ref{fig_norm_ber} displays the normalized BER of the
proposed scheme (\ref{eq_prob1a}-\ref{eq_mk}, \ref{eq
eta_correlation}) as a function of the eigenvalues of the Markov
transition matrices ($\lambda_{1}=1$ and $\lambda_{2}$) for $L=100$.
The BER is normalized by $P_{b}^{Kab}$, the estimated BER if no local bias
modification is applied in (\ref{eq eta_correlation}) (i.e., the
ordinary MUD of Kabashima is used (\ref{eq eta})). A substantial
improvement in the BER performance due to the proposed scheme is
presented. Notice that for very high eigenvalues ($\lambda_{1}=1,
\lambda_{2}>0.995$) the normalized BER grows since for these
eigenvalues $\xi\geq L$ (e.g., $\xi (\lambda_{2}=0.995) \approx
200>L=100$ ), which leads to strong effects of wrong detected bits
over the detected word. Figure~\ref{fig_norm_ber} also presents the
normalized BER under realistic model of a mismatch in the generating
Markov transition matrix estimation. Considering a symmetric Markov
transition matrix, a mismatch of $\pm10 \%$ of element $T_{-1-1}$ is
examined. That is to say, the detector assumes that the probability
to detect $\hat{b}_{k}^{l}=-1, \hat{b}_{k}^{l-1}=-1$ is higher/lower
than the true probability. Clearly, the proposed scheme still
suggests a substantial improvement in the BER for these practical
settings. The sudden growing of the BER for the positive mismatch
when $\lambda_{2}>0.8$ is due to the detrimental effect of infinite
correlation length.

As stated in the introduction, the mainstream alternative to our
approach is source coding, or compression~\cite{BookCoverThomas}. In
order to evaluate the attractiveness of the proposed scheme, we
compare its bit error probability, $P_{b}^{\textrm{Corr}}$, to the
bit error probability, $P_{b}^{\textrm{Comp}}$, achieved by the
nearly optimal Kabashima's multiuser detection of the transmission
of unbiased (optimally) compressed source bits.
We consider the theoretical optimal compression bound without limiting ourselves to any concrete algorithm.
The ratio between these two probabilities is given by \BEA\label{eq_ratio}
\frac{P_{b}^{\textrm{Corr}}}{P_{b}^{\textrm{Comp}}}&&=\frac{P_{b}^{\textrm{Corr}}(T^{\textrm{Corr}},\sigma,\beta)}
{P_{b}^{\textrm{Comp}}(T^{\textrm{i.i.d}},\sigma,\beta^{\textrm{Comp}}=\beta
H_{b})} \nonumber\\&&=\frac{H_{b} \cdot
P_{b}^{\textrm{Corr}}(T^{\textrm{Corr}},\sigma,\beta)}
{P_{b}^{\textrm{Corr}}(T^{\textrm{i.i.d}},\sigma,\beta^{\textrm{Comp}}=\beta
H_{b})}, \EEA where $H_{b}=-\sum_{a,b=\pm1}\mu_{a}
T_{ab}^{\textrm{Corr}} \log_{2}(T_{ab}^{\textrm{Corr}})$ denotes the
binary source's entropy, $\mu_{a}$ denotes the stationary
distribution of the correlated source and $T^{\textrm{Corr}}$ and
$T^{\textrm{i.i.d}}$ denote the Markov transition matrix and
identically and independently distributed (i.i.d) matrix
respectively. Note that as compression results in $H_{b}$ times less
compressed bits, they can be transmitted using an $H_{b}$ times
lower the bandwidth per bit. In order to exploit the entire
bandwidth, the spreading codes should be expanded. We have assumed
that the compressed bits are conveyed under an effectively $H_{b}$
times lower load $\beta$. Also, in computing $P_{b}^{\textrm{Comp}}$
when we assume an optimal source code, asymptotically speaking, a
single error in detecting a compressed bit leads, on average, to
$1/H_{b}$ errors in the uncompressed information.

\begin{figure}
\begin{center}
\begin{tabular}{c c}
{\scalebox{1}[1]{\includegraphics[width=0.67\textwidth]{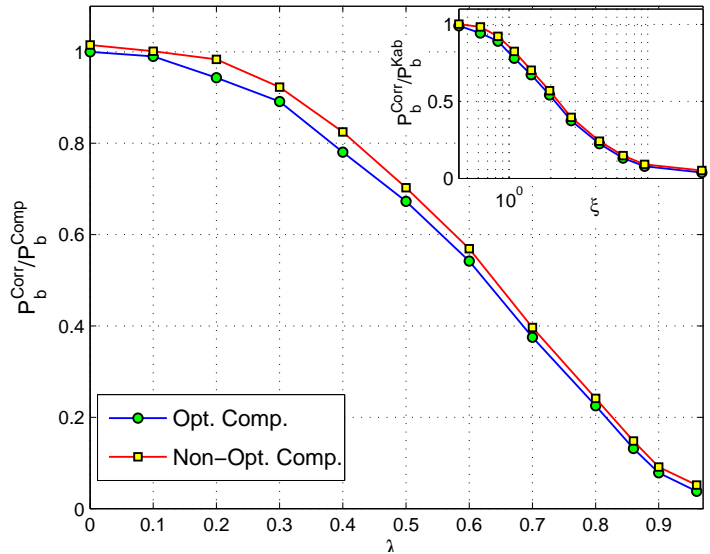}}}
\end{tabular}
\end{center}
\vspace{-0.4cm} \caption{A comparison of the BER obtained in
simulations, using the proposed scheme, $P_{b}^{Corr}$, versus
optimal and non-optimal (i.e., $5\%$ above the entropy) compression,
$P_{b}^{Comp}$. The ratio $\frac{P_{b}^{Corr}}{P_{b}^{Comp}}$ is
plotted as a function of the $2^{nd}$ eigenvalue of the Markov
transition matrix, $\lambda_{2}$, for load $\beta=0.8$, $\sigma=0.8$
and $L=30$. The comparison of the BER as a function of the correlation length $\xi$ is drawn in the inset.}\label{fig_Markov_vs_compression}
\end{figure}

Figure~\ref{fig_Markov_vs_compression} presents simulation results
of the BER ratio~(\ref{eq_ratio}) as a function of the eigenvalues
of the Markov transition matrices ($\lambda_{1}=1, \lambda_{2}$).
Interestingly, applying the proposed scheme is superior to the
optimal compression alternative for all $\lambda_{2}$ of the
Markov transition matrices.

Figure~\ref{fig_Markov_vs_compression} also demonstrates the
increasing superiority of the proposed scheme over non-optimal
compression. A sub-optimal fixed-length source encoder compressing
at $5\%$ above entropy is assumed. Again, this curve is obtained
using~(\ref{eq_ratio}), where the optimal compression rate $H_{b}$
is now replaced by the sub-optimal rate $(1+0.05)H_{b}$.

Note that the same trend of the ratio $\frac{P_{b}^{Corr}}{P_{b}^{Comp}}$ is also valid for $\beta>1$ (for example, $\beta=1.5$).

Another possible way to compare the proposed scheme and the
alternative of source coding, is to keep the parameters $\beta$ and
$\sigma$ constant in both approaches. While the BER of the proposed
scheme is calculated as explained above, the BER performance of the
source coding is calculated as following.
Applying an optimal encoder/decoder scheme, we can encode a correlated sequence into the same size as for the uncompressed case using the rate
$Rate = H_{b}$. After using the the MUD algorithm of of Kabashima for i.i.d. sequence we end up with bit error rate, $P_{b}^{Kab}$, which is now the starting point for the optimal decoder of a binary symmetric channel (BSC) with flip rate $f=P_{b}^{Kab}$ (we assume error bits are uncorrelated). The residue error, $P_{b}^{Comp}$, after the  separation scheme, the MUD of Kabashima and the optimal decoder is given by the following equation
\BE \label{H_b}
H_{b}=\frac{1-H_{2}(f)}{1-H_{2}\left(P_{b}^{\textrm{Comp}}\right)}.
\EE
Typical results are given in the following table.\\\\
\begin{tabular}{||l|l|l|l|l||} \hline
$\sigma $ & $\beta$ &$P_{b}^{\textrm{Corr}}$&$P_{b}^{\textrm{Comp}}$
& $P_{b}^{\textrm{Corr}}/P_{b}^{\textrm{Comp}} $\\ \hline \hline
$0.8$ & $0.8$ &  $0.034$ &  $0.042$ &  $0.81$\\ \hline
$1$ & $1$ &  $0.067$ &  $0.111$ &  $0.6$\\ \hline
\end{tabular}
\\\\Note that superiority of the proposed scheme relative to the
optimal compression alternative is still valid.

\section{Summary}
We have introduced a new scheme for detecting correlated sources
directly, with no need for source coding, by using a message-passing
based multiuser detector. The detection was applied simultaneously
over a block of transmitted binary symbols. The BER improvement for
a transmitted block while using the suggested method, have been
demonstrated. We have exhibited simulation results which
demonstrated a substantial BER improvement in comparison with the
unmodified detector and the alternative of source compression. The
robustness of our scheme have been shown, under practical model
settings, including wrong estimation of the generating Markov
transition matrix at the receiver and when using finite-length
spreading codes.

An important factor in implementing such a scheme is its
computational cost. The complexity of Kabashima's iterative detector
\cite{Kabashima}, when detecting a bit for a single user is
$\mathcal{O}(tN)$, where $t$ is the number of iterations required
for convergence. Adapting it for the case of $K$ users, each
transmits a word of length $L$, the complexity is $\mathcal{O}(L t K
N)$. In the proposed scheme, where all the words of all the users
are decoded simultaneously, the complexity is $\mathcal{O}(L
\tilde{t} K N)$. Simulations indicate that $t$ and $\tilde{t}$ are
comparable, although applying the exist algorithm \cite{Kabashima}
will result in a higher BER. The proposed scheme manages to incorporate the knowledge of the correlations between the transmitted binary symbols in the existing algorithm \cite{Kabashima} without increasing computational cost. Though the proposed scheme is heuristic, it still obtains better BER performance relative to the unmodified detector and the alternative of source compression.

Another important factor is the updating scheme of the applied local bias calculation on each binary symbol over the transmitted block of information. Four updating schemes were tested on the proposed detector: Parallel updating scheme (PUS), sequential updating scheme (SUS), back-front updating scheme (BFUS) and random-sequential updating scheme (RSUS). Simulations indicate that the best BER performance and convergence rate, in iterations, is achieved while using SUS or BFUS.

\end{document}